# Attosecond pulses at kiloelectronvolt photon energies from high-order harmonic generation with core electrons

Christian Buth,[1, 2, *] Feng He (何峰),[2, 3] Joachim Ullrich,[2, 4] Christoph H. Keitel,[2] and Karen Z. Hatsagortsyan[2]

[1]*Argonne National Laboratory, Argonne, Illinois 60439, USA*
[2]*Max-Planck-Institut für Kernphysik, Saupfercheckweg 1, 69117 Heidelberg, Germany*
[3]*Laboratory for Laser Plasmas and Department of Physics,
Shanghai Jiao Tong University, Shanghai 200240, China*
[4]*Physikalisch-Technische Bundesanstalt, Bundesallee 100, 38116 Braunschweig, Germany*
(Dated: 23 Jul 2014)

High-order harmonic generation (HHG) in simultaneous intense near-infrared (NIR) laser light and brilliant x rays above an inner-shell absorption edge is examined. A tightly bound inner-shell electron is transferred into the continuum. Then, NIR light takes over and drives the liberated electron through the continuum until it eventually returns to the cation leading in some cases to recombination and emission of a high-order harmonic (HH) photon that is upshifted by the x-ray photon energy. We develop a theory of this scenario and apply it to 1s electrons of neon atoms. The boosted HH light is used to generate a single attosecond pulse in the kiloelectronvolt regime. Prospects for nonlinear x-ray physics and HHG-based spectroscopy involving core orbitals are discussed.

## I. INTRODUCTION

High-order harmonic generation (HHG) by atoms in intense near-infrared (NIR) laser fields is a fascinating phenomenon and a versatile tool; it has spawned the field of attoscience, is used for spectroscopy, and serves as a light source in many optical laboratories [1–3]. Within the single-active electron (SAE) approximation [4] and the three-step model of HHG [5], the NIR laser tunnel ionizes a valence electron and accelerates it in the continuum. When the NIR laser field changes direction, the liberated electron is driven back to rescatter (or recollide) with the parent ion. This may cause the electron to recombine with the ion whereby the excess energy due to the atomic potential and due to the energy gained from the NIR laser field is released in terms of a high-order harmonic (HH) of the NIR frequency. In HHG each rescattered wave packet generates an individual attosecond burst of radiation. Thus using an ultrashort NIR laser pulse leads to an attosecond pulse train [6] in which the individual bursts are separated by a half cycle period of the NIR laser from which a single burst can be isolated by filtering out the harmonics close to the HHG cutoff [7].

The recollision-scenario-based HHG is efficient up to a driving laser intensity of $\sim 10^{16}\,\mathrm{W/cm^2}$ at infrared wavelengths which corresponds to a HH photon energy at the cutoff of $\sim 10\,\mathrm{keV}$. At higher intensities, the relativistic drift motion of the continuum electron in the direction of propagation of the driving laser field prohibits recollision and, consequently, suppresses HHG [8]. There are methods to counteract the relativistic drift [3, 9, 10], nonetheless, the HH photon yield decreases dramatically with rising photon energy due to the reduced recombination cross section and the decreased probabilities of ionization per harmonic. In the nonrelativistic regime, HHG has been pushed to its limits in recent works using mid-infrared lasers [11] at intensities of $\sim 10^{15}\,\mathrm{W/cm^2}$. Long-wavelength driving lasers allow a large ponderomotive potential at relatively low electric field strengths which decreases the ionization background and improves phase-matching of the HH light in a macroscopic medium with respect to shorter driving laser wavelengths. Meanwhile, the low probability of HH emission at high photon energies are compensated by using high gas pressures [12].

Present-day theory of HHG mostly gravitates around HHG from valence electrons and the SAE approximation [1–3, 5, 13]. Frequently, this mindset is also applied to two-color HHG where a NIR laser is combined with VUV/XUV light which, thereby, assists in the ionization process leading to an overall increased yield [14]. This principle is evolved further by using attosecond XUV pulses to modify the HHG process which increases the yield for a certain frequency range by enhancing the contribution from specific quantum orbits [15]. However, there are only few exceptions, e.g., Refs. 16, 17, in which many-electron effects are treated for two-color HHG. Fleischer and Moiseyev [16] include implicitly other electrons by using a frequency-dependent polarizability for the atoms. Then the XUV light is found to cause new plateaus to emerge at higher energies, however, with a much lower HH yield. Explicit two-electron effects in two-color HHG are considered in Refs. 17; there, the XUV photon energy is tuned to the core-valence resonance in the transient cation that is produced in the course of the HHG process by tunnel ionization. This leads to a second high-yield plateau that is shifted to higher energies by the XUV photon energy. A precursor of HHG boosted by x rays is x-ray and optical wave mixing—specifically sum-frequency generation (SFG) of x rays and an optical laser—which was demonstrated at LCLS by Glover *et al.* [18] with a conversion efficiency of $3 \times 10^{-7}$. This experiment offers a novel way to probe matter and is a very encouraging motivation of our theoretical research on x-ray boosted HHG, to be discussed in what follows. X-ray and optical wave mixing was proposed theoretically [19] in the 1970s but only recently it has been observed experimentally due to the novel FELs with unprecedented x-ray intensity [18]. The successful SFG of NIR light and x raysys inspires confidence in the prospects of x-ray-boosted HHG.

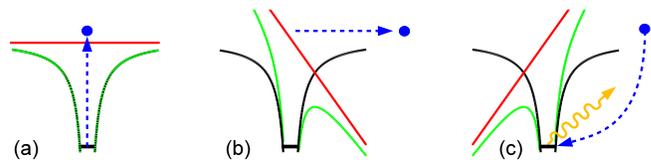

FIG. 1. (Color online) Schematic of the modified three-step model of HHG with core electrons ionized by x rays: (a) x-ray absorption ejects a core electron (b) which is subsequently driven through the continuum by the NIR light; (c) upon returning to the parent ion it may recombine with the core hole and release its excess energy in terms of a high-order harmonic of the NIR laser upshifted by the x-ray photon energy. The NIR laser has only a noticeable influence on valence and continuum electrons and barely influences core electrons. The converse holds true for the interaction with the x rays.

---

* Corresponding author. Present address: Theoretische Chemie, Physikalisch-Chemisches Institut, Ruprecht-Karls-Universität Heidelberg, Im Neuenheimer Feld 229, 69120 Heidelberg, Germany. Electronic mail; christian.buth@web.de

In this paper, we consider x-ray photons from a free electron laser (FEL) such as the Linac Coherent Light Source (LCLS) [20] that couple core electrons directly to the continuum. Our SAE scheme is depicted in Fig. 1 and proceeds in allusion to the three-step model of HHG [5] as follows: (a) the atomic core is ionized by one-x-ray-photon absorption; (b) the liberated electron propagates freely in the electric field of the NIR laser; (c) the direction of the NIR field is reversed and in some cases the electron is driven back to the ion and recombines with it emitting HH radiation. Although Fig. 1a suggest that the electron is born at a specific phase of the NIR light, this is not the case; instead, electrons are ejected during the entire x-ray pulse. Because of that electrons appear in the continuum at any phase of the NIR laser which is in contrast to the usual NIR-laser-only HHG [1] where ionization predominantly occurs close to the peak of the electric field of the NIR laser. We exemplify our method for $1s$ core electrons of neon with an ionization potential (IP) of $I_\text{P} = 870.2\,\text{eV}$ [21]. This yields HH radiation close to kiloelectronvolt photon energies from which we isolate a single attosecond pulse.

X-ray-boosted HHG offers completely new prospects for HH spectroscopy by involving core electrons; it thus goes beyond a mere extension of the HHG cutoff into the kiloelectronvolt regime for which one may use conventional concepts that are laboratory size and valence-electron based. The paper is structured as follows. In Sec. II we develop a theory of x-ray boosted HHG with core electrons. Results of single-atom HHG and estimates of phase matching in a macroscopic medium are discussed in Sec. III. We conclude our work in Sec. IV.

All equations are formulated in atomic units [22]. The Bohr radius $1\,\text{bohr} = 1\,a_0$ is the unit of length, $1\,t_0$ represents the unit of time, the unit of energy is $1\,\text{hartree} = 1\,\text{E}_h$, and charge is quantified in units of the elementary charge $1\,e$. Intensities are given in units of $1\,\text{E}_h\,t_0^{-1}\,a_0^{-2} = 6.43641 \times 10^{15}\,\text{W}\,\text{cm}^{-2}$. Electric polarizabilities are specified in $1\,e^2\,a_0^2\,\text{E}_h^{-1} = 1.64878 \times 10^{-41}\,\text{C}^2\,\text{m}^2\,\text{J}^{-1}$.

## II. THEORY

In the following, we develop a quantum theory of x-ray ionization-based HHG using the SAE [4]. This approximation is justified by recent works on neon excited and ionized by x rays which show that the excitation and ionization process by x rays is well-described by the independent electron approximation [23–28]. However, electron correlation in the ground state and final states play a role in the photoionization process, particularly in the near-threshold regime, and need to be accounted for, if the accuracy of the description is to be improved. The ground state $|0\rangle$ is represented by the electron in the core orbital of the atomic ground state. Continuum states are described by plane waves $|\vec{k}\rangle$ for $\vec{k} \in \mathbb{R}^3$ where we neglect the impact of the atomic potential which is termed the strong-field approximation [13, 29]. The Hamiltonian of the model is

$$\hat{H} = \hat{H}_\text{A} + \hat{H}_\text{L} + \hat{H}_\text{X} \,, \tag{1}$$

with the atomic electronic structure Hamiltonian $\hat{H}_\text{A}$ and the interaction of the SAE with the NIR laser $\hat{H}_\text{L}$ and the x rays $\hat{H}_\text{X}$ is expressed in electric dipole approximation in length form [30]. The SAE-model electronic structure is

$$\hat{H}_\text{A} = |0\rangle \left(-I_\text{P} - \frac{\text{i}}{2}\Gamma'_0(t)\right)\langle 0| \tag{2}$$
$$+ \int_{\mathbb{R}^3} |\vec{k}\rangle \left(\frac{\vec{k}^2}{2} - \frac{\text{i}}{2}\Gamma_{\vec{k}}(t)\right)\langle \vec{k}|\,\text{d}^3k \,,$$

where $I_\text{P}$ is the IP of the core electron. In Eq. (2), we need to account for the finite lifetime of the involved states due to ionization and Auger decay. The total destruction rate of the ground state due to x-ray and NIR-laser-induced ionization—that involves also core electrons—is expressed as $\Gamma_0(t)$ whereas $\Gamma'_0(t)$ stands for ionization of valence electrons only. The destruction of core-ionized states by Auger decay [31] and ionization is given by $\Gamma_{\vec{k}}(t)$. Here, $\Gamma'_0(t)$, $\Gamma_0(t)$, and $\Gamma_{\vec{k}}(t)$ depend on time because photoionization is determined by the envelope of the NIR and x-ray pulses. The subscript "$\vec{k}$" is only a label as $\Gamma_{\vec{k}}(t)$ depends negligibly on $\vec{k}$ for fast continuum electrons. For a linearly polarized NIR laser with an electric field $\vec{E}_{\mathrm{L}}(t)$, we have

$$\hat{H}_{\mathrm{L}} = \int_{\mathbb{R}^3} \int_{\mathbb{R}^3} |\vec{k}\rangle \langle \vec{k}| \vec{E}_{\mathrm{L}}(t) \cdot \vec{r} |\vec{k}'\rangle \langle \vec{k}'| \, \mathrm{d}^3k \, \mathrm{d}^3k' , \quad (3)$$

where $\vec{r}$ is the electric dipole operator. The form of $\hat{H}_{\mathrm{L}}$ incorporates only the impact of the NIR laser on continuum electrons; it does not describe the coupling of valence electrons to the continuum. Ionization is treated via $\Gamma'_0(t)$, $\Gamma_0(t)$, and $\Gamma_{\vec{k}}(t)$ whereas conventional HHG from valence electrons is not included; it can be described via the established approach of Lewenstein et al. [13]. Further, the interaction with the x rays is given by

$$\hat{H}_{\mathrm{X}} = \int_{\mathbb{R}^3} |\vec{k}\rangle \langle \vec{k}| E_{\mathrm{X}}(t) \vec{e}_{\mathrm{X}} \cdot \vec{r} |0\rangle \langle 0| \, \mathrm{d}^3k + \mathrm{h.c.} , \quad (4)$$

with the electric field $E_{\mathrm{X}}(t)$ and the linear polarization vector $\vec{e}_{\mathrm{X}}$ of the x rays. Inner-shell electrons are tightly bound such that the NIR laser hardly affects them. Hence, in the transition matrix elements only the x-ray term is present, i.e., $\langle \vec{k}| \hat{H}_{\mathrm{X}} |0\rangle$.

Based on the above model of an atom in two-color light, we derive equations of motion (EOMs) for its time evolution using the following ansatz [13] for the SAE wave packet

$$|\Psi, t\rangle = \mathrm{e}^{\mathrm{i}\, I_{\mathrm{P}}\, t} \left[ a(t) |0\rangle + \int_{\mathbb{R}^3} b(\vec{k}, t) |\vec{k}\rangle \, \mathrm{d}^3k \right] . \quad (5)$$

Inserting Eq. (5) into the time-dependent Schrödinger equation $\mathrm{i} \frac{\partial}{\partial t} |\Psi, t\rangle = \hat{H} |\Psi, t\rangle$ and projecting onto $\langle 0|$ yields the EOM for the ground-state amplitude

$$\mathrm{i}\, \dot{a}(t) = -\frac{\mathrm{i}}{2} \Gamma'_0(t)\, a(t) + E_{\mathrm{X}}(t) \int_{\mathbb{R}^3} b(\vec{k}, t) \langle 0| \vec{e}_{\mathrm{X}} \cdot \vec{r} |\vec{k}\rangle \, \mathrm{d}^3k . \quad (6)$$

Projecting onto $\langle \vec{k}|$ for all $\vec{k} \in \mathbb{R}^3$ results in EOMs for the continuum amplitudes

$$\mathrm{i}\, \frac{\partial b(\vec{k}, t)}{\partial t} = \left[ I_{\mathrm{P}} + \frac{\vec{k}^2}{2} - \frac{\mathrm{i}}{2} \Gamma_{\vec{k}}(t) \right] b(\vec{k}, t) + \mathrm{i}\, \vec{E}_{\mathrm{L}}(t) \cdot \vec{\nabla}_k b(\vec{k}, t) + a(t)\, E_{\mathrm{X}}(t)\, \langle \vec{k}| \vec{e}_{\mathrm{X}} \cdot \vec{r} |0\rangle . \quad (7)$$

The NIR laser related matrix elements are determined using Eq. (3) yielding $\langle \vec{k}| \hat{H}_{\mathrm{L}} |\vec{k}'\rangle = \mathrm{i}\vec{E}_{\mathrm{L}}(t) \cdot \vec{\nabla}_k \delta^3(\vec{k}-\vec{k}')$.

To solve the coupled system of linear first-order partial integrodifferential equations (6) and (7), we realize that the second term on the right-hand side of Eq. (6) represents the core-ionization rate by the x rays. Hence we may drop this term and account for its contribution by replacing $\Gamma'_0(t)$ by $\Gamma_0(t)$. This approximation decouples Eq. (6) from Eq. (7) which can now be integrated assuming that there is no light for $t < 0$. The solution is

$$a(t) = \begin{cases} 1 & ; t < 0 \\ \mathrm{e}^{-\frac{1}{2} F_0(t)} & ; t \geq 0 , \end{cases} \quad (8)$$

where the temporal destruction exponent is defined by

$$F_i(t) = \theta(t) \int_0^t \Gamma_i(t')\, \mathrm{d}t' , \quad (9)$$

with the Heaviside step function $\theta$ [32] for $i \in \{0\} \cup \{\vec{k} \mid \forall \vec{k} \in \mathbb{R}^3\}$. With the closed-form solution of Eq. (8), we integrate Eq. (7) formally [13] by introducing the canonical momentum $\vec{p} = \vec{k} - \vec{A}_{\mathrm{L}}(t)$ and the vector potential of the NIR laser $\vec{A}_{\mathrm{L}}(t) = -\int_{-\infty}^{t} \vec{E}_{\mathrm{L}}(t')\, \mathrm{d}t'$. This yields for the continuum amplitudes

$$\begin{aligned} b(\vec{k}, t) = (-\mathrm{i}) \int_0^t & a(t')\, E_{\mathrm{X}}(t') \langle \vec{p} + \vec{A}_{\mathrm{L}}(t')| \vec{e}_{\mathrm{X}} \cdot \vec{r} |0\rangle \\ & \times \mathrm{e}^{-\mathrm{i}\, S(\vec{p}, t, t')}\, \mathrm{e}^{-\mathrm{i}\, \omega_{\mathrm{X}}\, (t-t')} \\ & \times \mathrm{e}^{-\frac{1}{2}(F_{\vec{k}}(t) - F_{\vec{k}}(t'))}\, \mathrm{d}t' \, \Big|_{\vec{p}=\vec{k}-\vec{A}_{\mathrm{L}}(t)} , \end{aligned} \quad (10)$$

with the quasiclassical action

$$S(\vec{p}, t, t') = \int_{t'}^{t} \left[ \frac{1}{2} \left( \vec{p} + \vec{A}_{\mathrm{L}}(t'') \right)^2 + I_{\mathrm{P}} - \omega_{\mathrm{X}} \right] \mathrm{d}t'' , \quad (11)$$

where we neglected the dependence of $F_{\vec{k}}(t)$ on $\vec{k}$ in Eq. (10).

Having solved the EOMs, we calculate the HH emission along the direction $\vec{e}_{\mathrm{D}}$ which is determined by the electric dipole transition matrix element [5, 13, 17, 33]:

$$D(t) = \int_{\mathbb{R}^3} a^*(t) \langle 0| \vec{e}_{\mathrm{D}} \cdot \vec{r} |\vec{k}\rangle\, b(\vec{k}, t)\, \mathrm{d}^3k = D'(t)\, \mathrm{e}^{-\mathrm{i}\, \omega_{\mathrm{X}}\, t} , \quad (12)$$

introducing the slowly-varying dipole moment $D'(t)$. Inserting Eqs. (8), (10) and (11) into Eq. (12) and using the saddle-point approximation [32] to simplify the triple integration over $\vec{k}$ produces

$$\begin{aligned} D'(t) = -\frac{\mathrm{i}}{2} \int_0^t & \sqrt{\frac{(-2\pi\mathrm{i})^3}{\tau^3}}\, \langle 0| \vec{e}_{\mathrm{D}} \cdot \vec{r} |\vec{p}_{\mathrm{st}}(t,\tau) + \vec{A}_{\mathrm{L}}(t)\rangle \\ & \times \langle \vec{p}_{\mathrm{st}}(t,\tau) + \vec{A}_{\mathrm{L}}(t-\tau) | \vec{e}_{\mathrm{X}} \cdot \vec{r} |0\rangle\, \mathrm{e}^{-\mathrm{i}\, S_{\mathrm{st}}(t,\tau)} \\ & \times \mathcal{E}_{\mathrm{X}}(t-\tau)\, \mathrm{e}^{-\frac{1}{2}[F_0(t)+F_0(t-\tau)+F_{\vec{k}}(t)-F_{\vec{k}}(t-\tau)]}\, \mathrm{d}\tau . \end{aligned} \quad (13)$$

Here we introduced the excursion time $\tau = t - t'$ of the electron in the continuum; at the stationary point (saddle point), for which $\vec{\nabla}_p S(\vec{p}, t, t') = 0$ holds, the canonical momentum is

$$\vec{p}_{\mathrm{st}}(t, \tau) = -\frac{1}{\tau} \int_{t-\tau}^{t} \vec{A}_{\mathrm{L}}(t')\, \mathrm{d}t' \quad (14)$$

and the quasiclassical action [Eq. (11)] at the stationary point is $S_{\mathrm{st}}(t, \tau) \equiv S(\vec{p}_{\mathrm{st}}(t, \tau), t, t - \tau)$. The term $\mathrm{e}^{-\mathrm{i}\,\omega_\mathrm{X}\,(t-t')}$ in the middle line of Eq. (10) was omitted by employing the rotating-wave approximation [30], i.e., only the positive frequency components of the x-ray electric field $E_\mathrm{X}^+(t)$ remain. Specifically, we introduce the decomposition

$$E_\mathrm{X}^+(t) = \frac{1}{2}\,\mathcal{E}_\mathrm{X}(t)\,\mathrm{e}^{-\mathrm{i}\,\omega_\mathrm{X}\,t} \qquad (15)$$

with the complex field envelope $\mathcal{E}_\mathrm{X}(t) = E_{0\mathrm{X}}(t)\,\mathrm{e}^{-\mathrm{i}\,[\varphi_\mathrm{X}(t)+\varphi_{\mathrm{X},0}]}$, the real field envelope $E_{0\mathrm{X}}(t)$, the time-dependent phase $\varphi_\mathrm{X}(t)$, and the carrier to envelope phase (CEP) $\varphi_{\mathrm{X},0}$ [34].

The emission of HH light is quantified by the harmonic photon number spectrum (HPNS) [17, 33] of a single atom which is the probability density function [32] to find a HH photon with specific energy $\omega$. Along the propagation axis of the NIR laser, the HPNS is

$$\frac{\mathrm{d}^2 P(\omega)}{\mathrm{d}\omega\,\mathrm{d}\Omega} = 4\,\pi\,\omega\,\varrho(\omega)\,|\tilde{D}(\omega)|^2\;, \qquad (16)$$

with the density of free-photon states $\varrho(\omega) = \frac{\omega^2}{(2\pi)^3\,c^3}$ in the interval $[\omega, \omega + \mathrm{d}\omega]$ [30], the speed of light in vacuum $c$ and the solid angle $\Omega$. The HHG spectrum of Eq. (16) follows from the Fourier transformation [32] of $D(t)$ denoted by $\tilde{D}(\omega)$; the transforms of $D(t)$ and $D'(t)$ [Eq. (12)] are related by $\tilde{D}(\omega) = \tilde{D}'(\omega - \omega_\mathrm{X})$, i.e., the entire harmonic spectrum is shifted by $\omega_\mathrm{X}$ toward higher energies.

## III. RESULTS AND DISCUSSION

### A. Single-atom high-order harmonic generation

We apply our theory to generate HH radiation from $1s$ core electrons of a neon atom where the linear polarization vectors of NIR laser $\vec{e}_\mathrm{L}$, x rays $\vec{e}_\mathrm{X}$, and HH emission $\vec{e}_\mathrm{D}$ are along the $z$ axis. The atomic orbitals of neutral neon are calculated in Hartree-Fock-Slater approximation [36] with the program of Herman and Skillman setting $\alpha = 1$ [37]; they are used to evaluate the electric dipole matrix elements in Eq. (13) following Sec. V A of Ref. 38. Present-day FELs generate x rays on the basis of the self-amplification of spontaneous emission (SASE) principle [20]. We model SASE pulses with the partial coherence method [28, 39]. Two sample pulses are displayed in Fig. 2 where modeling parameters are specified in the figure caption. The chosen x-ray photon energy $\omega_\mathrm{X} = 881.8\,\mathrm{eV}$ is above the $K$ edge but still fairly close to it such that ionization by x rays is efficient. The rate of destruction of neutral and core-ionized neon is determined as follows. For destruction with photoionization by x rays, we use the total ionization cross section (involving $1s$, $2s$, and $2p$ electrons) at $\omega_\mathrm{X}$ which is $\sigma_0 = 3.0 \times 10^{-19}\,\mathrm{cm}^2$ for the neutral atom and $\sigma_{\vec{k}} = 3.1 \times 10^{-20}\,\mathrm{cm}^2$ for the core-ionized cation obtained with Ref. 40. Double core hole formation via ionization of the remaining core electron of the core-ionized state is energetically not possible for the chosen $\omega_\mathrm{X}$; core-excitation of a second electron is also not included in $\sigma_{\vec{k}}$. Additionally, the NIR laser destroys the system by tunnel ionization which occurs with the instantaneous rate $\Gamma_{0,\mathrm{L}}(t)$ that is determined with the Ammosov-Delone-Krainov (ADK) formula [41] and the valence IP of neon $21.5645\,\mathrm{eV}$ [42]. At the chosen peak intensity [caption of Fig. 3], the cycle-averaged destruction rate of a neutral neon atom at the peak intensity of the NIR laser is $0.03\,\mathrm{eV}$. The ionization rate for a core-ionized neon cation is set to zero, i.e., $\Gamma_{\vec{k},\mathrm{L}}(t) = 0$, because the valence IP of core-ionized states is much larger than the valence IP of the neutral atom leading to a vanishingly small width due to the exponential dependence of the ADK rate on the IP [41]. Summing up all destructive contributions yields for a neutral neon atom

$$\Gamma_0(t) = \sigma_0\,J_\mathrm{X}(t) + \Gamma_{0,\mathrm{L}}(t) \qquad (17)$$

and for a core-ionized neon cation

$$\Gamma_{\vec{k}}(t) = \Gamma_\mathrm{c} + \sigma_{\vec{k}}\,J_\mathrm{X}(t) + \Gamma_{\vec{k},\mathrm{L}}(t)\;, \qquad (18)$$

where $J_\mathrm{X}(t)$ is the instantaneous x-ray flux. In Eq. (18), we account also for destruction of the system by Auger decay of the core hole [31]. The experimental width of a Ne $1s$ core hole, $\Gamma_\mathrm{c} = 0.27\,\mathrm{eV}$, is taken from Ref. 21.

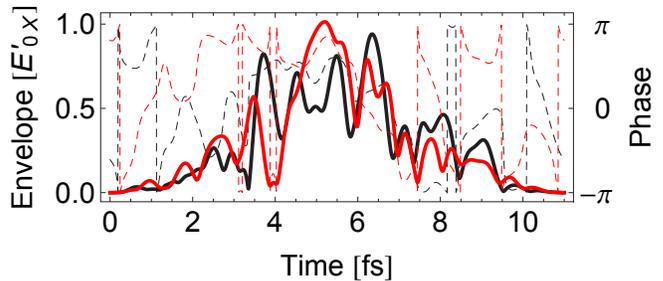

FIG. 2. (Color) Two exemplary SASE x-ray FEL pulses (black and red) for an x-ray photon energy of $\omega_\mathrm{X} = 569\,\omega_\mathrm{L} = 881.8\,\mathrm{eV}$. The electric field envelope $E_{0\mathrm{X}}(t)$ is specified in units of the peak electric field $E'_{0\mathrm{X}}$ for the peak intensity of $2.2 \times 10^{16}\,\mathrm{W/cm}^2$ and is represented by solid lines whereas the phase $\varphi_\mathrm{X}(t)$ is depicted as dashed lines with CEP $\varphi_{\mathrm{X},0} = 0$. The average FWHM pulse duration of an ensemble of SASE pulses is 1.5 optical cycles with a cosine square pulse envelope [35]. The average pulse spectrum is taken to be Gaussian with a FWHM bandwidth of $8\,\mathrm{eV}$ which corresponds to a coherence time of $0.24\,\mathrm{fs}$.

Although neon atoms are destroyed by the mechanisms discussed above and are thus no longer available for the HHG process that is modeled in this paper, there are secondary HHG processes involving the produced cations. Namely, HHG spectra from neon dications which are produced by Auger decay are shifted significantly in energy with respect to the spectra discussed here because the returning continuum electron can only recombine with a valence orbital and consequently the added energy difference due to the IP for valence orbitals is much smaller than for core orbitals. Therefore, these HHG processes are excluded from our treatment; their light can be suppressed experimentally by a suitable filter [43]. Further, for two-color light, additional photoionization occurs with fairly low probability for neutral and core-ionized neon induced by both, the NIR laser and the x rays. As we choose the x-ray energy quite close to the core-ionization threshold, destruction by double core ionization, etc. is energetically forbidden [26] and only core excitations and valence ionization by x rays are allowed which are fairly small. The resulting HHG spectra from the produced neon dications are again shifted noticeably in energy with respect to the spectra discussed here because of the changed IP and thus can be clearly distinguished.

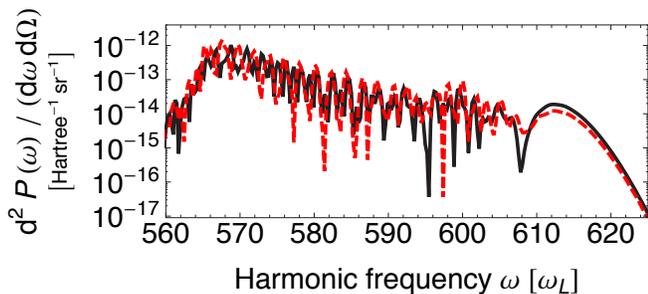

FIG. 3. (Color online) Harmonic photon number spectra (HPNS) [Eq. (16)] for $1s$ core electrons of neon using the SASE pulses of Fig. 2. The solid, black line is from the black SASE pulse and the dashed, red line from the red pulse. The x-ray intensity [caption of Fig. 2] was chosen such that the ionization probability is the same as the tunnel ionization probability by the NIR laser. The NIR laser vector potential $\tilde{A}_\mathrm{L}(t)$ has a cosine square pulse envelope [35] with a FWHM duration of 1.5 optical cycles, CEP $\varphi_{\mathrm{L},0} = \pi/2$, and a peak intensity of $5 \times 10^{14}\,\mathrm{W/cm^2}$ for 800 nm wavelength, i.e., the photon energy is $\omega_\mathrm{L} = 1.55\,\mathrm{eV}$.

Spectra from HHG with $1s$ core electrons of neon for the two SASE pulses of Fig. 2 are shown in Fig. 3. We see an upshifted HPNS toward higher harmonic orders as well as a appreciable impact of the SASE pulse shape on the spectra as was also found for x-ray boosted HHG-based on resonant core excitations of the transient cation during the HHG process [17]. Although a variation due to SASE pulses is clearly visible, it is not dramatic and the overall structure of the two HPNS is similar. For an accurate description of the HPNS below the IP of the core shell of $I_\mathrm{P} = 870.2\,\mathrm{eV} = 861\,\omega_\mathrm{L}$ [21], also influences from the Rydberg states of neon need to be taken into account (e.g., Refs. 44 and References therein) which were not included in our theory of Sec. II. Hence these parts of the HPNS are not shown in Fig. 3.

To gain deeper insight into the temporal evolution of the emission of HH light from core electrons, we carry out a time-frequency analysis of the dipole moment $D_i(t)$ [Eq. (12)] for SASE pulse $i \in \{1,2\}$ from Fig. 2 (the first and the second pulses are drawn in black and red, respectively) with a short-time Fourier transform—also called windowed Fourier transform—which reads

$$\check{\tilde{D}}_i(t,\omega) = \int_{-\infty}^{\infty} D_i(\tau)\,W_i(\tau - t)\,\mathrm{e}^{\mathrm{i}\,\omega\,\tau}\,\mathrm{d}\tau, \qquad (19)$$

using a Gaussian window function

$$W_i(t) = \frac{1}{\sqrt{2\pi}\,\sigma_\mathrm{w}}\,\mathrm{e}^{-\frac{t^2}{2\sigma_\mathrm{w}^2}}, \qquad (20)$$

with a variance of $\sigma_\mathrm{w} = 0.1\,\mathrm{fs}$ where the transform is not particularly sensitive to this choice. The resulting time and frequency dependent dipole moment $\check{\tilde{D}}_i(t,\omega)$ is employed in a difference plot in Fig. 4 that shows the function

$$\Delta(t,\omega) = 10^{-\left|\lg|\check{\tilde{D}}_1(t,\omega)|^2 - \lg|\check{\tilde{D}}_2(t,\omega)|^2\right|}. \qquad (21)$$

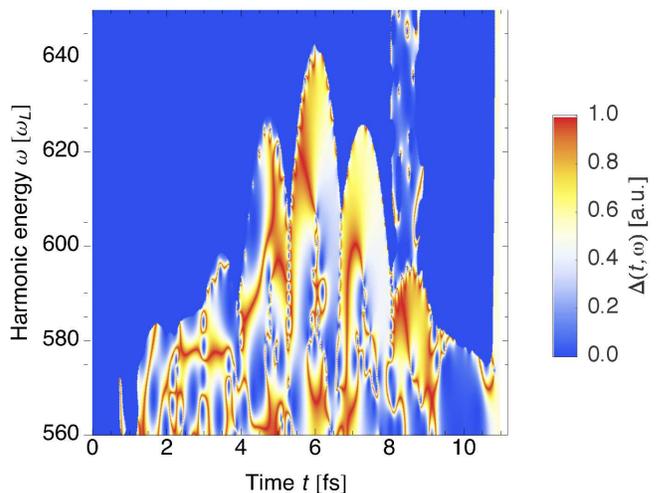

FIG. 4. (Color) Difference plot [Eq. (21)] of the time-frequency analysis [Eq. 19] of HHG with $1s$ core electrons of neon for the black and the red SASE pulses of Fig. 2. The plotted values are unity for $|\check{\tilde{D}}_1(t,\omega)| = |\check{\tilde{D}}_2(t,\omega)|$ and they are zero in the limit that the absolute value of the difference between both dipole moments goes to infinity. For the plotting, we set values $\lg|\check{\tilde{D}}_1(t,\omega)|^2 < -22$ to $-80$ and values $\lg|\check{\tilde{D}}_2(t,\omega)|^2 < -22$ to $-40$ in Eq. (21).

For all $t, \omega$ for which the absolute values of the windowed Fourier transforms of the two pulses are equal, $\Delta(t,\omega)$ is unity. Conversely, the larger the difference between the windowed Fourier transforms, the closer $\Delta(t,\omega)$ approaches zero.

Inspecting the difference plot [Eq. (21)] in Fig. 4 of the time-frequency analysis of the dipole moment [Eq. (19)] for the two FEL x-ray pulses of Fig. 2, reveals that the imprinting of the SASE pulse shape onto the HPNS is noticeable but the temporal evolution of HH emission is largely determined by the NIR laser. Clearly, the cycles of the NIR laser can be identified and the emission of HH radiation from x-ray-boosted HHG follows it closely. For the harmonics close to the cutoff, only the probability of harmonic emission differs between the HPNS for the two FEL x-ray pulses preserving the pattern imposed by the NIR laser. The largest influence of the SASE pulse shape is on the low harmonics up to a harmonic order of 600. Here the cycle structure of the NIR laser is significantly altered in the difference plot and largely dissolved in a flame-like shape indicating an appreciable impact of the SASE pulse shape onto HH emission. Namely, for NIR-laser-only HHG, the highest HH photon energies are generated from electrons that are ionized close to the maximum of the NIR laser field [1]. Inspecting Fig. 4, we find a similar property for x-ray-boosted HHG where the highest harmonic order are emitted around the maximum of the electric field. This implies that these harmonics are generated during a short time interval in which only a small part of the FEL x-ray pulse influences their generation causing only a variation of the HH yield but not a noticeable modification of the shape of the HPNS close to the cutoff. Conversely, low harmonic orders are generated by longer stretches of the NIR laser pulse, i.e., longer time intervals of the FEL x-ray pulse are sampled. This leads to a much larger dependence of the lower harmonics on the FEL x-ray pulse shape as is revealed in Fig. 4.

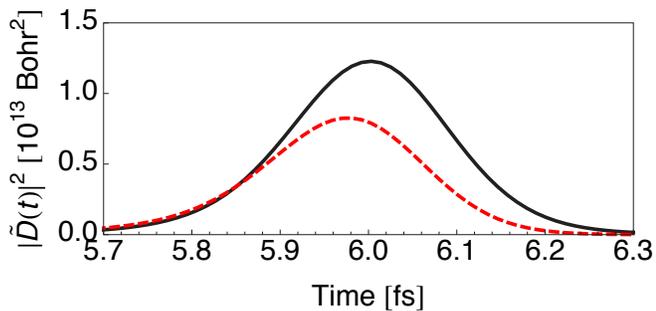

FIG. 5. (Color online) Attosecond pulses from x-ray-boosted HHG for harmonics with $\omega \geq 608\,\omega_{\rm L}$. The pulse shapes are approximated by Gaussian functions with a FWHM duration of 220 as. Line styles as in Fig. 3.

For the NIR-laser field of 1.5-cycle FWHM duration used to compute Fig. 4, the intensity of each cycle is different; the highest harmonic orders are produced at the peak of the NIR laser pulse where the single maximum of the vector potential produces the orders larger than 608. By filtering out harmonic orders lower than 608, we obtain the single attosecond pulses shown in Fig. 5 for the two FEL x-ray pulses of Fig. 2 that we fit with a Gaussian profile of FWHM duration of about 220 as. The SASE pulse shape impacts only the peak intensity of the attosecond pulses.

The difference between NIR laser-only HHG from Ne $2p$ valence electrons and x-ray boosted HHG with Ne $1s$ core electrons, is caused by the following four points. First, the initial ionization step is mediated, on the one hand, by tunnel ionization and, on the other hand, by a one-x-ray-photon absorption process. Tunnel ionization occurs predominantly along the linear NIR laser polarization axis leading to a continuum electron wave packet elongated along this axis [41]. In contrast, one-x-ray-photon ionization of a Ne $1s$ electron produces a wave packet with the figure-8-shape of a $p$ wave [38]. Hence only part of the electron wave packet which is ejected along $\vec{e}_{\rm L}$ will have a significant probability to recombine with the ionic remnant and thus contribute to HHG. Second, core-hole recombination is less efficient than valence-hole recombination due to more compact core orbitals compared with valence orbitals. This causes the dipole matrix element, which is responsible for core-hole recombination, to have a smaller overall magnitude than for valence-hole recombination. Third, valence holes are stable whereas core holes decay. Hence the HH yield for core holes is reduced compared with valence holes because there is a significant probability for core holes to decay during the excursion of the core electron in the continuum. Fourth, tunnel ionization releases electrons at rest but one-x-ray-photon ionization ejects electrons with a kinetic energy of $\omega_{\rm X} - I_{\rm P}$. For $\omega_{\rm X} \gg I_{\rm P}$, the NIR laser electric field may not be strong enough to drive liberated electrons back to the atom such that they cannot recombine with the parent ion and no HH light is emitted.

### B. High-order harmonic generation in a macroscopic medium

We have investigated x-ray boosted HHG from a single atom. In order to predict the yield of HHG from a macroscopic ensemble of atoms, one needs to examine the propagation of the FEL x rays, the NIR laser light, and the generated HH light in a dense gas [Fig. 6a], i.e., solve the Maxwell equations together with the Schrödinger equation [45]. This task is out of the scope of the present paper and, instead, we provide a simple estimate of the order of magnitude of the macroscopic yield, in the following.

As a macroscopic medium, we use a cylindrical gas cell filled with neon gas of an atom number density of $n_{\rm a} = 10^{18}$ cm$^{-3}$ and a length of $L = 0.1$ mm. The gas cell is placed within the Rayleigh range $z_{\rm R} = \pi \frac{w_0^2}{\lambda_{\rm L}} = 2.3$ mm, i.e., $L \leq 2\,z_{\rm R}$, of the Gaussian NIR-laser beam [34] for a beam waist of $w_0 = 30\,\lambda_{\rm L}$ and a wavelength in vacuum of $\lambda_{\rm L} = \frac{2\pi c}{\omega_{\rm L}} = 800$ nm. The gas cell has a radius of $w_0$ such that the interaction volume becomes $V = \pi\,w_0^2\,L$. In order to achieve a substantial output of x-ray-boosted HH radiation, the chosen length $L$ of the medium needs to be smaller than the absorption length of the FEL x rays and the x-ray boosted HH light which is given by the inverse of the absorption coefficient $\mu = \sigma_0\,n_{\rm a}$ [31, 46]. This yields $\mu^{-1} = 3.5$ cm which is much larger than the length of the gas cell. Thus the condition $\mu^{-1} < L$ is well fulfilled.

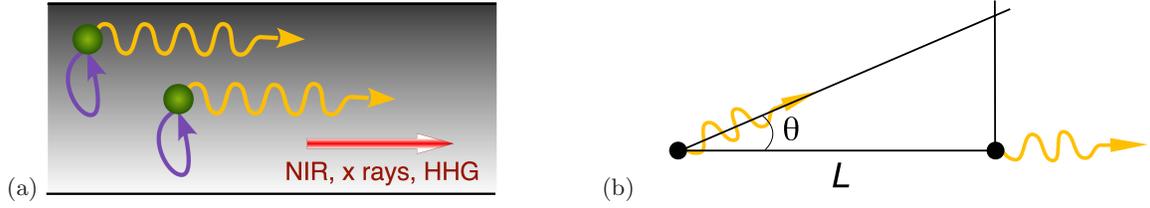

FIG. 6. (Color online) Schematic of (a) propagation of a NIR laser pulse, FEL x rays, and the generated HH light in a macroscopic medium. (b) Longitudinal phase matching condition of HH emission by two atoms. Here, $L$ is the length of the macroscopic medium and $\theta$ is the opening angle of the cone of coherent HH emission.

The index of refraction of the propagation of NIR light, the FEL x rays and the generated HH x rays in the medium is determined following the Appendix of Ref. 46. The largest of these indices is the refraction of the NIR light which is given by

$$n_L = n_{L,a} + n_{L,p} - 1 \; ; \tag{22}$$

it consists of the refraction due to the polarizability of the atoms $n_{L,a}$ and the refraction due to free electrons from ionization of the medium $n_{L,p}$. The contribution $n_{L,a}$ is obtained via the dynamic polarizability of neon atoms for $\omega_L = 1.55\,\text{eV}$ which is $\alpha_L = 2.7075\,e^2\,a_0^2\,E_h^{-1}$ [47] and follows with

$$n_{L,a} = 1 + 2\pi n_a \alpha_L \tag{23}$$

to $n_{L,a} - 1 = 2.5 \times 10^{-6}$ [46]. The free electrons add a negative term $n_{L,p} - 1$ to the refractive index $n_L$, where the corresponding index of refraction is

$$n_{L,p} = \sqrt{1 - \frac{\omega_p^2}{\omega_L^2}} \tag{24}$$

and the plasma frequency is $\omega_p = \sqrt{4\pi n_e}$ [46, 48] with the free electron density $n_e$. The free electrons arise due to ionization by the NIR laser and the FEL x rays. The probability of ionization of an atom in the gas by the NIR laser is given by the integral over the pulse duration of the instantaneous ADK rate [41] for tunnel ionization. For the pulse parameters of the NIR laser from the caption of Fig. 3, a probability of $P = 0.083$ to ionize an atom in the medium is obtained. The ensemble-averaged ionization probability from FEL x-ray pulses is chosen to be the same as $P$ which leads to the parameters in the caption of Fig. 2. Depending on the actual shape of a specific SASE pulse, an x-ray-induced ionization probability is found that slightly differs from $P$ which is, however, neglected in what follows. The number density of free electrons after irradiation by the NIR laser and FEL x rays is $n_e = 3\,P\,n_a = 2.5 \times 10^{17}\,\text{cm}^{-3}$. The factor of 3 accounts for the fact that half the atoms are ionized by FEL x rays leading predominantly to core holes which decay in neon almost exclusively by Auger decay producing dications. Production of higher charge states via ionization by the NIR light and x rays and ionization by the HH light are neglected. The index of refraction due to free electrons is $n_{L,p} - 1 \approx -\frac{\omega_p^2}{2\omega_L^2} = -7.2 \times 10^{-5}$ and the total refractive index of the NIR light becomes $n_L - 1 = -6.9 \times 10^{-5}$.

The refractive index for the x rays from the FEL and the HHG is generally very small [23, 24, 31, 46]. Although the FEL x-ray photon energy of $\omega_X = 881.8\,\text{eV}$ is quite close to the ionization potential of neon $1s$ electrons of $I_P = 870.2\,\text{eV}$ [21], resonant effects are already small in this region [23, 24, 46] and the formula

$$n_R(\omega) = 1 - \frac{2\pi Z n_a}{\omega^2} \;, \tag{25}$$

can be used where $Z = 10$ is the number of electrons in a neon atom [31]. For the FEL x rays, we obtain $n_X - 1 = n_R(\omega_X) - 1 = -8.9 \times 10^{-9}$ and, accordingly, for the HH light $n_H - 1 = n_R(\omega_H) - 1 = -7.5 \times 10^{-9}$ with a photon energy $\omega_H = h\,\omega_L = 960.9\,\text{eV}$ close to the x-ray-boosted cutoff at harmonic order $h = 620$.

As NIR-laser-only HHG depends only on the phase of the NIR laser, there is a stable phase relation between the emitted HH light from spatially separated atoms. This is mostly also the case for x-ray boosted HHG [Fig. 4] although the impact of the FEL x rays plays a role. In order to estimate phase matching in a macroscopic medium, we consider longitudinally separated atoms in a macroscopic medium in Fig. 6b. The HH light adds coherently, when the phase difference between HH emission between the spatially separated atoms is smaller than $\pi$. However, SASE FELs have only a limited longitudinal (temporal) coherence [20] as is seen by the phase fluctuations in Fig. 2. As can be seen in Fig. 4, HH emission is not too much impacted by fluctuating phases of the FEL x rays but largely dominated by the phase of the NIR laser. Therefore, we assume for our estimate here that there is approximately no impact of the fluctuating phase of SASE pulses on the longitudinal coherence of HH emission. Moreover, the NIR and x-ray pulses have a total longitudinal extend of only $3.3\,\mu\text{m}$ which is shorter than the medium length $L$. Clearly, these are issues that need to be investigated further with pulse propagation [45].

Now let us analyze the phase-matching condition. The harmonic emission phase driven by the FEL x rays and the NIR laser depends on the coordinate of the emitter

atom $\vec{r}$ via

$$\phi(\vec{r}) = \vec{k} \cdot \vec{r} = \left(\frac{\omega_{\rm H}}{\omega_{\rm L}} \vec{k}_{\rm L} - \vec{k}_{\rm H} + \vec{k}_{\rm X} - \frac{\omega_{\rm X}}{\omega_{\rm L}} \vec{k}_{\rm L}\right) \cdot \vec{r}, \quad (26)$$

see Eq. (32) of Ref. 9. The phase difference of the harmonic field emitted from two atoms at points $\vec{r}_1 = x_1\,\vec{e}_x$ and $\vec{r}_2 = x_2\,\vec{e}_x$ along the propagation axis of the NIR laser and the FEL x rays is

$$\Delta\phi = \phi(\vec{r}_2) - \phi(\vec{r}_1) = \vec{k} \cdot (\vec{r}_2 - \vec{r}_1). \quad (27)$$

The HH field propagates along the $x$ axis with a small angular divergence $\theta \ll 1$ as depicted in Fig. 6b giving rise to a HH phase difference of

$$|\vec{k}_{\rm H} \cdot (\vec{r}_2 - \vec{r}_1)| = |\vec{k}_{\rm H}|\,\Delta x\,\cos\theta \approx n_{\rm H}\,\frac{\omega_{\rm H}}{c}\,\Delta x\left(1 - \frac{\theta^2}{2}\right), \quad (28)$$

where $\Delta x = x_2 - x_1$. Then the phase-matching condition becomes

$$|\Delta\phi| \approx \left|\frac{\omega_{\rm H}\,\Delta x}{c}\left[n_{\rm H}\,\frac{\theta^2}{2} + (n_{\rm L} - n_{\rm H}) - \frac{\omega_{\rm X}}{\omega_{\rm H}}(n_{\rm L} - n_{\rm X})\right]\right| < \pi. \quad (29)$$

It is expressed as follows

$$\left|n_{\rm H}\,\frac{\theta^2}{2} + (n_{\rm L} - 1)\left(1 - \frac{\omega_{\rm X}}{h\,\omega_{\rm L}}\right)\right| < \frac{\lambda_{\rm L}}{2\,h\,L}, \quad (30)$$

for the highest harmonic $h$ and the length of the medium $L$. Using the assumption that the $\theta$-dependent term, which is always nonnegative, is larger than the magnitude of the other term on the left hand side, if that one is negative, relation (30) simplifies to

$$\theta < \sqrt{\frac{2}{n_{\rm H}}\left[\frac{\lambda_{\rm L}}{2\,h\,L} - (n_{\rm L} - 1)\left(1 - \frac{\omega_{\rm X}}{h\,\omega_{\rm L}}\right)\right]}. \quad (31)$$

The coherence length of the medium quantifies the maximum length of the medium that still leads to coherent emission. It is derived from Eq. (30) by setting $\theta = 0$, replacing $L$ by $\ell_{\rm coh}$, and turning the relation into an equation which leads to

$$\ell_{\rm coh} = \frac{\lambda_{\rm L}}{2\,h\,\left|(n_{\rm L} - 1)\left(1 - \frac{\omega_{\rm X}}{h\,\omega_{\rm L}}\right)\right|}, \quad (32)$$

i.e., the medium length should be restricted by $L \leq \ell_{\rm coh} = 0.11\,{\rm mm}$.

To estimate the total harmonic yield, we use relation (31) to determine the angular spread of harmonics to $\theta = 4.9 \times 10^{-3}\,{\rm rad}$. Within this cone centered on an atom, another atoms in the cone emits HH light coherently with the other atom. The cone has an opening angle of $\theta$ along the propagation axis and subtends a solid angle $\Delta\Omega$ of

$$\Delta\Omega = \int_0^{2\pi}\int_0^{\theta}\sin\vartheta\,{\rm d}\vartheta\,{\rm d}\varphi \approx \pi\,\theta^2. \quad (33)$$

Then, the HH photon number emitted per pulse by the medium is found from the HH spectrum of a single atom [Eq. (16)] by

$$N_{\rm ph} = (n_{\rm a}\,V)^2\,\Delta\Omega \int_{\omega_{\rm X}}^{\infty}\frac{{\rm d}^2 P(\omega)}{{\rm d}\omega\,{\rm d}\Omega}\,{\rm d}\omega, \quad (34)$$

where the factor $(n_{\rm a}\,V)^2$ is caused by the coherent emission by the atoms in the medium. The $N_{\rm ph}$ is estimated to be $\sim 10^7$ photons which is a sufficiently high to make x-ray boosted HH light interesting for applications. In the attosecond pulses, there are $\sim 10^5$ photons. Furthermore, $N_{\rm ph}$ can be increased by orders of magnitude using a higher x-ray intensity and—if one is only interested in a large HH yield—longer NIR- and x-ray light pulses. Using a suitably chosen material filter, one can absorb low energy x rays and thus isolate the attosecond pulse, e.g., mylar with a thickness of $10\,\mu{\rm m}$ at a density of $1.4\,\frac{\rm g}{{\rm cm}^3}$ has a rapidly increasing transmission around $1000\,{\rm eV}$ [43].

## IV. CONCLUSION

Using x rays to ionize core electrons as a first step in HHG, we predict high-order harmonics boosted into the x-ray regime. By superimposing the harmonics close to the cutoff, we have obtained single attosecond pulses. Our approach complements FEL-based strategies to reduce the pulse duration of x rays—which are presently several femtoseconds—to attoseconds (Ref. 49 and References therein). It has the advantage that the attosecond x-ray pulses have a defined phase-relation to the NIR laser and that x-ray-boosted HHG can be employed at any FEL with moderate cost and minimal impact on other experiments.

The focus of this paper is to model the basic physics underlying HHG with core electrons ionized by x rays without aiming for high accuracy of the HH yield. Therefore, we base the theory on the three-step model [5] and the strong-field approximation [13, 29]. Within these approximations, the obtained HH yield is somewhat overestimated by up to an order of magnitude. Using the Eikonal-Volkov approximation (EVA) [50] or numerical integration of the time-dependent Schrödinger equation in SAE approximation [51] are routes to obtain a more accurate HH yield.

Our scheme extends most HHG-based methods to inner-shell atomic and molecular orbitals using suitably tuned x rays. Specifically, we see a potential for the following applications. The HHG spectra depend on the pulse shape of SASE x rays; a reconstruction with frequency-resolved optical gating (FROG) [52] may be possible—thus offering the long-sought pulse characterization for SASE light [53]—but requires further theoretical investigation. Finally, the emitted upshifted HH light due to core recombination bears the signature of the *core* orbital; thus it can be used for ultrafast time-dependent chemical imaging [54] involving inner shells which is not feasible so far.

## ACKNOWLEDGMENTS

We would like to thank Stefano M. Cavaletto, Marcelo F. Ciappina, Markus C. Kohler, and Ji-Cai Liu (刘纪彩) for helpful discussions. C.B. was supported by the Chemical Sciences, Geosciences, and Biosciences Division of the Office of Basic Energy Sciences, Office of Science, U.S. Department of Energy, under Contract No. DE-AC02-06CH11357 and by a Marie Curie International Reintegration Grant within the 7$^{\text{th}}$ European Community Framework Program (call identifier: FP7-PEOPLE-2010-RG, proposal No. 266551). F.H. was supported by the Pujiang scholar fellowship (grant No. 11PJ1404800), the National Science Foundation of China (grant Nos. 11175120 and 11104180), the National Science Foundation in Shanghai (grant No. 11ZR1417100), and the Fok Ying-Tong Education Foundation for Young Teachers in the High Education Institutions of China (grant No. 131010).

[1] P. Agostini and L. F. DiMauro, Rep. Prog. Phys. **67**, 813 (2004).
[2] F. Krausz and M. Ivanov, Rev. Mod. Phys. **81**, 163 (2009).
[3] M. C. Kohler, T. Pfeifer, K. Z. Hatsagortsyan, and C. H. Keitel, Adv. At. Mol. Opt. Phys. **61**, 159 (2012), arXiv:1201.5094.
[4] K. C. Kulander and T. N. Rescigno, Comput. Phys. Commun. **63**, 523 (1991); K. C. Kulander, K. J. Schafer, and J. L. Krause, in *Super-Intense Laser-Atom Physics*, NATO Advanced Study Institute Series B: Physics, Vol. 316, edited by B. Piraux, A. L'Huillier, and K. Rzążewski (Plenum Press, New York, 1993) pp. 95–110.
[5] K. J. Schafer, B. Yang, L. F. DiMauro, and K. C. Kulander, Phys. Rev. Lett. **70**, 1599 (1993); P. B. Corkum, Phys. Rev. Lett. **71**, 1994 (1993).
[6] P. M. Paul, E. S. Toma, P. Breger, G. Mullot, F. Augé, P. Balcou, H. G. Muller, and P. Agostini, Science **292**, 1689 (2001).
[7] E. Goulielmakis, M. Schultze, M. Hofstetter, V. S. Yakovlev, J. Gagnon, M. Uiberacker, A. L. Aquila, E. M. Gullikson, D. T. Attwood, R. Kienberger, F. Krausz, and U. Kleineberg, Science **320**, 1614 (2008).
[8] A. Di Piazza, C. Müller, K. Z. Hatsagortsyan, and C. H. Keitel, Rev. Mod. Phys. **84**, 1177 (2012).
[9] M. C. Kohler and K. Z. Hatsagortsyan, Phys. Rev. A **85**, 023819 (2012).
[10] M. C. Kohler, M. Klaiber, K. Z. Hatsagortsyan, and C. H. Keitel, EPL (Europhysics Letters) **94**, 14002 (2011).
[11] J. Tate, T. Auguste, H. G. Muller, P. Salières, P. Agostini, and L. F. DiMauro, Phys. Rev. Lett. **98**, 013901 (2007).
[12] T. Popmintchev, M.-C. Chen, A. Bahabad, M. Gerrity, P. Sidorenko, O. Cohen, I. P. Christov, M. M. Murnane, and H. C. Kapteyn, Proc. Natl. Acad. Sci. U.S.A. **106**, 10516 (2009); P. Arpin, T. Popmintchev, N. L. Wagner, A. L. Lytle, O. Cohen, H. C. Kapteyn, and M. M. Murnane, Phys. Rev. Lett. **103**, 143901 (2009); M.-C. Chen, P. Arpin, T. Popmintchev, M. Gerrity, B. Zhang, M. Seaberg, D. Popmintchev, M. M. Murnane, and H. C. Kapteyn, Phys. Rev. Lett. **105**, 173901 (2010); T. Popmintchev, M.-C. Chen, D. Popmintchev, P. Arpin, S. Brown, S. Ališauskas, G. Andriukaitis, T. Balčiunas, O. D. Mücke, A. Pugzlys, A. Baltuška, B. Shim, S. E. Schrauth, A. Gaeta, C. Hernández-García, L. Plaja, A. Becker, A. Jaron-Becker, M. M. Murnane, and H. C. Kapteyn, Science **336**, 1287 (2012).
[13] M. Lewenstein, P. Balcou, M. Y. Ivanov, A. L'Huillier, and P. B. Corkum, Phys. Rev. A **49**, 2117 (1994).
[14] K. Ishikawa, Phys. Rev. Lett. **91**, 043002 (2003); E. J. Takahashi, T. Kanai, K. L. Ishikawa, Y. Nabekawa, and K. Midorikawa, Phys. Rev. Lett. **99**, 053904 (2007); K. L. Ishikawa, E. J. Takahashi, and K. Midorikawa, Phys. Rev. A **80**, 011807 (2009); S. V. Popruzhenko, D. F. Zaretsky, and W. Becker, Phys. Rev. A **81**, 063417 (2010).
[15] K. J. Schafer, M. B. Gaarde, A. Heinrich, J. Biegert, and U. Keller, Phys. Rev. Lett. **92**, 023003 (2004); M. B. Gaarde, K. J. Schafer, A. Heinrich, J. Biegert, and U. Keller, Phys. Rev. A **72**, 013411 (2005); C. Figueira de Morisson Faria, P. Salières, P. Villain, and M. Lewenstein, Phys. Rev. A **74**, 053416 (2006); A. Heinrich, W. Kornelis, M. P. Anscombe, C. P. Hauri, P. Schlup, J. Biegert, and U. Keller, J. Phys. B **39**, S275 (2006); J. Biegert, A. Heinrich, C. P. Hauri, W. Kornelis, P. Schlup, M. P. Anscombe, M. B. Gaarde, K. J. Schafer, and U. Keller, J. Mod. Opt. **53**, 87 (2006); C. Figueira de Morisson Faria and P. Salières, Las. Phys. **17**, 390 (2007).
[16] A. Fleischer and N. Moiseyev, Phys. Rev. A **77**, 010102(R) (2008); A. Fleischer, Phys. Rev. A **78**, 053413 (2008).
[17] C. Buth, M. C. Kohler, J. Ullrich, and C. H. Keitel, Opt. Lett. **36**, 3530 (2011), arXiv:1012.4930; M. C. Kohler, C. Müller, C. Buth, A. B. Voitkiv, K. Z. Hatsagortsyan, J. Ullrich, T. Pfeifer, and C. H. Keitel, in *Multiphoton Processes and Attosecond Physics*, Springer Proceedings in Physics, Vol. 125, edited by K. Yamanouchi and K. Midorikawa (Springer, Berlin, Heidelberg, 2012) pp. 209–217, Conference proceedings of ICOMP12-ATTO3, arXiv:1111.3555; B. W. Adams, C. Buth, S. M. Cavaletto, J. Evers, Z. Harman, C. H. Keitel, A. Pálffy, A. Picón, R. Röhlsberger, Y. Rostovtsev, and K. Tamasaku, J. Mod. Opt. **60**, 2 (2013), Special Issue: Physics in Quantum Electronics. Selected Papers from the 42nd Winter Colloquium on the Physics of Quantum Electronics, 2–6 January 2012; C. Buth, submitted (2014), arXiv:1303.1332.
[18] T. E. Glover, D. M. Fritz, M. Cammarata, T. K. Allison, S. Coh, J. M. Feldkamp, H. Lemke, D. Zhu, Y. Feng, R. N. Coffee, M. Fuchs, S. Ghimire, J. Chen, S. Shwartz, D. A. Reis, S. E. Harris, and J. B. Hastings, Nature **488**,

603 (2012).

[19] I. Freund and B. F. Levine, Phys. Rev. Lett. **25**, 1241 (1970); Phys. Rev. Lett. **26**, 156 (1971); P. M. Eisenberger and S. L. McCall, Phys. Rev. A **3**, 1145 (1971).

[20] J. Arthur, P. Anfinrud, P. Audebert, K. Bane, I. Ben-Zvi, V. Bharadwaj, R. Bionta, P. Bolton, M. Borland, P. H. Bucksbaum, R. C. Cauble, J. Clendenin, M. Cornacchia, G. Decker, P. Den Hartog, S. Dierker, D. Dowell, D. Dungan, P. Emma, I. Evans, G. Faigel, R. Falcone, W. M. Fawley, M. Ferrario, A. S. Fisher, R. R. Freeman, J. Frisch, J. Galayda, J.-C. Gauthier, S. Gierman, E. Gluskin, W. Graves, J. Hajdu, J. Hastings, K. Hodgson, Z. Huang, R. Humphry, P. Ilinski, D. Imre, C. Jacobsen, C.-C. Kao, K. R. Kase, K.-J. Kim, R. Kirby, J. Kirz, L. Klaisner, P. Krejcik, K. Kulander, O. L. Landen, R. W. Lee, C. Lewis, C. Limborg, E. I. Lindau, A. Lumpkin, G. Materlik, S. Mao, J. Miao, S. Mochrie, E. Moog, S. Milton, G. Mulhollan, K. Nelson, W. R. Nelson, R. Neutze, A. Ng, D. Nguyen, H.-D. Nuhn, D. T. Palmer, J. M. Paterson, C. Pellegrini, S. Reiche, M. Renner, D. Riley, C. V. Robinson, S. H. Rokni, S. J. Rose, J. Rosenzweig, R. Ruland, G. Ruocco, D. Saenz, S. Sasaki, D. Sayre, J. Schmerge, D. Schneider, C. Schroeder, L. Serafini, F. Sette, S. Sinha, D. van der Spoel, B. Stephenson, G. Stupakov, M. Sutton, A. Szöke, R. Tatchyn, A. Toor, E. Trakhtenberg, I. Vasserman, N. Vinokurov, X. J. Wang, D. Waltz, J. S. Wark, E. Weckert, Wilson-Squire Group, H. Winick, M. Woodley, A. Wootton, M. Wulff, M. Xie, R. Yotam, L. Young, and A. Zewail, *Linac Coherent Light Source (LCLS): Conceptual Design Report*, Tech. Rep. SLAC-R-593, UC-414 (Stanford Linear Accelerator Center (SLAC), Menlo Park, California, USA, 2002); P. Emma, R. Akre, J. Arthur, R. Bionta, C. Bostedt, J. Bozek, A. Brachmann, P. Bucksbaum, R. Coffee, F.-J. Decker, Y. Ding, D. Dowell, S. Edstrom, J. Fisher, A. Frisch, S. Gilevich, J. Hastings, G. Hays, P. Hering, Z. Huang, R. Iverson, H. Loos, M. Messerschmidt, A. Miahnahri, S. Moeller, H.-D. Nuhn, G. Pile, D. Ratner, J. Rzepiela, D. Schultz, T. Smith, P. Stefan, H. Tompkins, J. Turner, J. Welch, W. White, J. Wu, G. Yocky, and J. Galayda, Nature Photon. **4**, 641 (2010).

[21] V. Schmidt, *Electron Spectrometry of Atoms Using Synchrotron Radiation* (Cambridge University Press, Cambridge, 1997).

[22] A. Szabo and N. S. Ostlund, *Modern Quantum Chemistry: Introduction to Advanced Electronic Structure Theory*, 1st, revised ed. (McGraw-Hill, New York, 1989).

[23] C. Buth, R. Santra, and L. Young, Phys. Rev. Lett. **98**, 253001 (2007), republished in Virtual Journal of Ultrafast Science **6**, issue 7 (2007), arXiv:0705.3615.

[24] C. Buth, R. Santra, and L. Young, Rev. Mex. Fís. S **56**, 59 (2010), Proceedings for the Pan American Advanced Studies Institute (PASI), Ultrafast and Ultrasmall; New Frontiers in AMO Physics, March 30 – April 11, 2008, Búzios, Brazil, arXiv:0805.2619.

[25] T. E. Glover, M. P. Hertlein, S. H. Southworth, T. K. Allison, J. van Tilborg, E. P. Kanter, B. Krässig, H. R. Varma, B. Rude, R. Santra, A. Belkacem, and L. Young, Nature Phys. **6**, 69 (2010).

[26] L. Young, E. P. Kanter, B. Krässig, Y. Li, A. M. March, S. T. Pratt, R. Santra, S. H. Southworth, N. Rohringer, L. F. DiMauro, G. Doumy, C. A. Roedig, N. Berrah, L. Fang, M. Hoener, P. H. Bucksbaum, J. P. Cryan, S. Ghimire, J. M. Glownia, D. A. Reis, J. D. Bozek, C. Bostedt, and M. Messerschmidt, Nature **466**, 56 (2010).

[27] E. P. Kanter, B. Krässig, Y. Li, A. M. March, P. Ho, N. Rohringer, R. Santra, S. H. Southworth, L. F. DiMauro, G. Doumy, C. A. Roedig, N. Berrah, L. Fang, M. Hoener, P. H. Bucksbaum, S. Ghimire, D. A. Reis, J. D. Bozek, C. Bostedt, M. Messerschmidt, and L. Young, Phys. Rev. Lett. **107**, 233001 (2011).

[28] S. M. Cavaletto, C. Buth, Z. Harman, E. P. Kanter, S. H. Southworth, L. Young, and C. H. Keitel, Phys. Rev. A **86**, 033402 (2012), arXiv:1205.4918.

[29] K. C. Kulander, Phys. Rev. A **38**, 778 (1988); G. G. Paulus, W. Nicklich, H. Xu, P. Lambropoulos, and H. Walther, Phys. Rev. Lett. **72**, 2851 (1994).

[30] M. O. Scully and M. S. Zubairy, *Quantum Optics* (Cambridge University Press, Cambridge, New York, Melbourne, 1997); P. Meystre and M. Sargent III, *Elements of Quantum Optics*, 3rd ed. (Springer, Berlin, 1999).

[31] J. Als-Nielsen and D. McMorrow, *Elements of Modern X-Ray Physics* (John Wiley & Sons, New York, 2001).

[32] G. B. Arfken and H. J. Weber, *Mathematical Methods for Physicists*, 6th ed. (Elsevier Academic Press, New York, 2005).

[33] D. J. Diestler, Phys. Rev. A **78**, 033814 (2008).

[34] J.-C. Diels and W. Rudolph, *Ultrashort Laser Pulse Phenomena*, 2nd ed., Optics and Photonics Series (Academic Press, Amsterdam, 2006).

[35] I. Barth and C. Lasser, J. Phys. B **42**, 235101 (2009).

[36] J. C. Slater, Phys. Rev. **81**, 385 (1951); J. C. Slater and K. H. Johnson, Phys. Rev. B **5**, 844 (1972).

[37] F. Herman and S. Skillman, *Atomic Structure Calculations* (Prentice-Hall, Englewood Cliffs, New Jersey, 1963); C. Buth and R. Santra, Phys. Rev. A **75**, 033412 (2007), republished in Virtual Journal of Ultrafast Science **6**, issue 4 (2007), arXiv:physics/0611122.

[38] C. Buth and K. J. Schafer, Phys. Rev. A **80**, 033410 (2009), republished in Virtual Journal of Ultrafast Science **8**, issue 10 (2009), arXiv:0905.3756.

[39] T. Pfeifer, Y. Jiang, S. Düsterer, R. Moshammer, and J. Ullrich, Opt. Lett. **35**, 3441 (2010); Y. H. Jiang, T. Pfeifer, A. Rudenko, O. Herrwerth, L. Foucar, M. Kurka, K. U. Kühnel, M. Lezius, M. F. Kling, X. Liu, K. Ueda, S. Düsterer, R. Treusch, C. D. Schröter, R. Moshammer, and J. Ullrich, Phys. Rev. A **82**, 041403(R) (2010).

[40] R. D. Cowan, *The Theory of Atomic Structure and Spectra*, Los Alamos Series in Basic and Applied Sciences (University of California Press, Berkeley, 1981); Los Alamos National Laboratory, Atomic Physics Codes, http://aphysics2.lanl.gov/tempweb/lanl.

[41] A. M. Perelomov, V. S. Popov, and V. M. Terentév, Zh. Exp. Theor. Fiz. **50**, 1393 (1966), [Sov. Phys. JETP **23**, 924–934 (1966)]; Zh. Exp. Theor. Fiz. **51**, 309 (1966), [Sov. Phys. JETP **24**, 207–217 (1967)]; A. M. Perelomov and V. S. Popov, Zh. Exp. Theor. Fiz. **52**, 514 (1967), [Sov. Phys. JETP **25**, 336–343 (1967)]; V. S. Popov, V. P. Kuznetsov, and A. M. Perelomov, Zh. Exp. Theor. Fiz. **53**, 331 (1967), [Sov. Phys. JETP **26**, 222–232 (1968)]; M. V. Ammosov, N. B. Delone, and V. P. Krainov, Zh. Eksp. Teor. Fiz. **91**, 2008 (1986), [Sov. Phys. JETP **64**, 1191–1194 (1986)]; N. B. Delone and V. P. Krainov, *Multiphoton Processes in Atoms*, 2nd ed., edited by P. Lambropoulos, Atoms and Plasmas, Vol. 13

(Springer, Berlin, 2000); G. L. Yudin and M. Y. Ivanov, Phys. Rev. A **64**, 013409 (2001).
[42] V. Kaufman and L. Minnhagen, J. Opt. Soc. Am. **62**, 92 (1972).
[43] B. L. Henke, E. M. Gullikson, and J. C. Davis, At. Data Nucl. Data Tables **54**, 181 (1993); "X-ray interactions with matter," The Center for X-Ray Optics, Lawrence Berkeley National Laboratory, Berkeley, California, USA (2010), http://henke.lbl.gov/optical_constants.
[44] S. X. Hu, A. F. Starace, W. Becker, W. Sandner, and D. B. Milošević, J. Phys. B **35**, 627 (2002); J.-C. Liu, M. C. Kohler, C. H. Keitel, and K. Z. Hatsagortsyan, Phys. Rev. A **84**, 063817 (2011).
[45] A. L'Huillier, K. J. Schafer, and K. C. Kulander, J. Phys. B **24**, 3315 (1991); E. Priori, G. Cerullo, M. Nisoli, S. Stagira, S. De Silvestri, P. Villoresi, L. Poletto, P. Ceccherini, C. Altucci, R. Bruzzese, and C. de Lisio, Phys. Rev. A **61**, 063801 (2000); M. B. Gaarde, J. L. Tate, and K. J. Schafer, J. Phys. B **41**, 132001 (2008).
[46] C. Buth and R. Santra, Phys. Rev. A **78**, 043409 (2008), republished in Virtual Journal of Ultrafast Science **7**, issue 11 (2008), arXiv:0809.3249.
[47] The dynamic polarizability of a neon atom at $\omega_{\rm L} = 1.55\,{\rm eV} = 0.057\,{\rm E}_h$ is determined by linear interpolation between the points $0.05\,{\rm E}_h$ and $0.1\,{\rm E}_h$ from Table 4, column "MCTDHF ab initio (exp.)" in Ref. 55.
[48] J. D. Jackson, *Classical Electrodynamics*, 3rd ed. (John Wiley & Sons, Chichester, New York, 1998).
[49] E. A. Schneidmiller and M. V. Yurkov, Phys. Rev. ST Accel. Beams **13**, 110701 (2010).
[50] O. Smirnova, M. Spanner, and M. Ivanov, Phys. Rev. A **77**, 033407 (2008).
[51] K. J. Schafer, in *Strong Field Laser Physics*, Springer Series in Optical Sciences, Vol. 134, edited by T. Brabec (Springer, New York, 2009) pp. 111–145.
[52] K. W. DeLong, D. N. Fittinghoff, R. Trebino, B. Kohler, and K. Wilson, Opt. Lett. **19**, 2152 (1994); R. Trebino, *Frequency-Resolved Optical Gating: The Measurement of Ultrashort Laser Pulses* (Kluwer Academic Publishers, Boston, Dordrecht, London, 2000).
[53] S. Düsterer, P. Radcliffe, C. Bostedt, J. Bozek, A. L. Cavalieri, R. Coffee, J. T. Costello, D. Cubaynes, L. F. DiMauro, Y. Ding, G. Doumy, F. Grüner, W. Helml, W. Schweinberger, R. Kienberger, A. R. Maier, M. Messerschmidt, V. Richardson, C. Roedig, T. Tschentscher, and M. Meyer, New J. Phys. **13**, 093024 (2011); M. Meyer, P. Radcliffe, T. Tschentscher, J. T. Costello, A. L. Cavalieri, I. Grguras, A. R. Maier, R. Kienberger, J. Bozek, C. Bostedt, S. Schorb, R. Coffee, M. Messerschmidt, C. Roedig, E. Sistrunk, L. F. DiMauro, G. Doumy, K. Ueda, S. Wada, S. Düsterer, A. K. Kazansky, and N. M. Kabachnik, Phys. Rev. Lett. **108**, 063007 (2012).
[54] T. Morishita, A.-T. Le, Z. Chen, and C. D. Lin, Phys. Rev. Lett. **100**, 013903 (2008).
[55] N. K. Rahman, A. Rizzo, and D. L. Yeager, Chem. Phys. Lett. **166**, 565 (1990).